\documentclass{emulateapj}

\usepackage{graphicx}
\usepackage{natbib}
\usepackage{txfonts}
\usepackage{color}
\usepackage{graphicx}

\def\lta{\lower2pt\hbox{$\buildrel {\scriptstyle <} 
   \over {\scriptstyle\sim}$}}
\def\gta{\lower2pt\hbox{$\buildrel {\scriptstyle >} 
   \over {\scriptstyle\sim}$}}

\begin{document}

\title{Simulations of GRB Jets in a Stratified External Medium: Dynamics, Afterglow Lightcurves, Jet Breaks and Radio Calorimetry}

\author{Fabio De Colle\altaffilmark{1}, Enrico Ramirez-Ruiz\altaffilmark{1},
Jonathan Granot\altaffilmark{2,3,4}, \& Diego Lopez-Camara\altaffilmark{5}}
\altaffiltext{1}{TASC, Department of Astronomy \& Astrophysics, University of California, Santa Cruz, CA 95064, USA; fabio@ucolick.org}
\altaffiltext{2}{Racah Institute of Physics, The Hebrew University, Jerusalem 91904, Israel}
\altaffiltext{3}{Raymond and Beverly Sackler School of Physics \& Astronomy, Tel Aviv University, Tel Aviv 69978, Israel}
\altaffiltext{4}{Centre for Astrophysics Research, University of Hertfordshire, College Lane, Hatfield, AL10 9AB, UK}
\altaffiltext{5}{Instituto de Ciencias Nucleares, Universidad Nacional Aut\'onoma de M\'exico, Ap. 70-543, 04510 D.F., M\'exico}


\begin{abstract}
The dynamics of gamma-ray burst (GRB) jets during the afterglow phase
is most reliably and accurately modeled using hydrodynamic
simulations. All published simulations so far, however, have
considered only a uniform external medium, while a stratified external
medium is expected around long duration GRB progenitors. Here we present simulations of the
dynamics of GRB jets and the resulting afterglow emission for both
uniform and stratified external media with $\rho_{\rm ext}
\propto r^{-k}$ for $k = 0,\,1,\,2$. The simulations are performed in
two dimensions using the special relativistic version of the {\it
Mezcal} code. Common to all calculations is the initiation of the GRB jet  as
a conical wedge of
half-opening angle $\theta_0 = 0.2$ whose radial profile is taken from the
self-similar Blandford-McKee solution. The dynamics for stratified
external media ($k = 1,\,2$) are broadly similar to those derived for  expansion into a
uniform external medium ($k=0$). The jet half-opening angle is observed to start
increasing logarithmically with time (or radius) once the Lorentz factor
$\Gamma$ drops below $\theta_0^{-1}$. For larger $k$ values, however, the lateral
expansion is faster at early times (when $\Gamma > \theta_0^{-1}$) and
slower at late times with  the jet expansion becoming Newtonian and slowly approaching
spherical symmetry over progressively longer timescales. We find that contrary to analytic expectations, there is
a reasonably sharp jet break in the lightcurve for $k = 2$ (a
wind-like external medium) although  the shape of the break is affected
more by the viewing angle (for $\theta_{\rm obs} \leq \theta_0$) than
by the slope of the external density profile (for $0\leq k\leq 2$). Steeper density profiles (i.e. increasing $k$ values) 
are found to produce  more gradual jet breaks while larger viewing angles cause 
smoother and later appearing jet breaks. The counter-jet becomes visible as it
becomes sub-relativistic, and for $k=0$ this results in a clear bump-like feature
in the light curve. However, for larger $k$ values the jet decelerates more
gradually,  causing only  a mild flattening in the radio  light curve that might be hard to discern
when  $k=2$. Late time radio calorimetry, which makes use of a spherical
flow approximation near the non-relativistic transition, is likely to consistently over-estimate the true
energy by up to a factor of a few for $k=2$, but either over-predict
or under-predict it by a smaller factor for $k = 0,1$. 
\end{abstract}




\section{Introduction}

The dynamics of gamma-ray burst (GRB) outflows depends on the density
distribution of the ambient medium as well as on the structure of the
relativistic expanding ejecta \citep[e.g.,][]{1998ApJ...499..301M}.
Up to the deceleration epoch, where most of the energy is transferred
to the shocked external medium, the dynamics is regulated by the
local radial structure of the ejecta, while at later times (as the
blastwave decelerates) it mainly depends on its global angular
structure. In the absence of characteristic scales, self-similar,
spherically symmetric solutions exist \citep[][hereafter  Blandford-McKee]{BM76}
and they are widely used to interpret observational data on GRB
afterglows.  However, even the simplest  departure from this
ideal model could drastically modify the afterglow
behavior. Anisotropies in the GRB outflow, for example, affect the
afterglow light curve when the mean jet energy per solid angle within
the visible region evolves significantly. As the jet decelerates, the
relativistic beaming weakens and the visible region increases.
If the outflow is collimated into a narrow jet with reasonably sharp
edges, this occurs at the time when the bulk Lorentz
factor $\Gamma$ equals the inverse of jet half-opening angle $\theta_0$. 
A simple analytic calculation using the usual scaling laws leads then to a steepening of
the afterglow flux decay rate, known as a {\it jet break}
~\citep[]{1997ApJ...487L...1R,SPH99,KP00}.  It is however clear from
numerical studies that such simple scalings do not provide an accurate
description of the
afterglow~\citep{Granot01,ZM09,2010A&A...520L...3M,vanEerten10a,WWF11,vEM11,
2011arXiv1110.5089V}.
Such numerical studies have so far been limited to the case
of a uniform external density while the interaction of relativistic
GRB jets with a non-uniform medium remains poorly understood.

Motivated by this, here we study the dynamics 
of two-dimensional (2D) axially symmetric impulsive jets
propagating in a spherically symmetric stratified medium of rest-mass
density $\rho=A r^{-k}$ and the resulting afterglow emission.
Since long duration GRBs  \citep{2009ARA&A..47..567G}  have massive star
progenitors whose winds are expected to modify their immediate
surroundings~\citep{2004ApJ...606..369C,2005ApJ...631..435R,2008A&A...478..769V,2011MNRAS.418..583M}, we
consider both steady and time varying stellar winds as possible
surrounding or external media for the GRB jet evolution.  The case
$k=2$ corresponds to a stellar wind for a massive star progenitor
~\citep{2000ApJ...536..195C,2000ApJ...543...66P,2001MNRAS.327..829R, 2005ApJ...619..968W}
with a constant ratio of its pre-explosion mass loss rate $\dot{M}_w$
and wind velocity $v_w$, in which case $\rho = A r^{-2}$, where $A
=\dot{M}_w/(4\pi v_w)$. However, since the dependence of $\dot{M}_w$
and $v_w$ on the time $t_w$ before the stellar explosion that triggers
the GRB is highly uncertain, it is worth considering other values of
$k$. For example, if $\dot{M}_w\propto t_w^a$ and $v_w\propto t_w^b$
then the location of a wind element at the time of the explosion
is $r = t_wv_w(t_w)\propto t_w^{1+b}$ so that $t_w\propto
r^{1/(1+b)}$ and we have $\dot{M}_w\propto r^{a/(1+b)}$, $v_w\propto
r^{b/(1+b)}$ and $\rho\propto r^{-2+(a-b)/(1+b)}$. For a constant
wind velocity ($b=0$) this gives $k = 2 - a$, which corresponds
to $k=2$ for $a=0$ (constant wind mass flux) and $k=1$ for $a=1$
(linearly increasing mass flux with time).

A brief description of our numerical methods and  initial
conditions for both  jet and external medium models is giving in
~\S~\ref{methods}.
 Detailed hydrodynamic simulations of GRB jets
interacting with $k=1,\,2$ stratified media are presented in
~\S~\ref{dynamics} and ~\S~\ref{rad}, where ~\S~\ref{dynamics} is
devoted to the jet dynamics and the resulting afterglow emission 
is discussed in ~\S~\ref{rad}.
 For completeness and comparison,
the interaction with a constant-density medium ($k=0$) is also
discussed, although the reader is referred to~\citet{paperI} for
a review of the current state of hydrodynamical modeling with
$k=0$. Our conclusions are summarized in ~\S~\ref{dis}.

\section{Numerical Methods}\label{methods}

\subsection{Code Description and Initial Conditions}

To study the dynamics of a GRB jet propagating in a stratified
external medium, we carry out a set of two-dimensional simulations
using the special relativistic hydrodynamic (SRHD) version of the
adaptive mesh refinement (AMR) code {\it Mezcal} \citep{paperI}.  The
{\it Mezcal} code integrates the SRHD equations by using a
second-order (in space and time, except in shocks where it reduces to
first order in space by a minmod limiter) upwind scheme based on the
relativistic HLL method \citep{1993JCoPh.105...92S}.  The equation of
state (EOS), relating enthalpy to pressure and density, is taken from
\citet{Ryu06}, which approximates the exact
\citet{1971tar..book.....S} EOS with an error of 0.5\%.  This EOS
properly recovers the correct values of the adiabatic index $\Gamma$
in the ultra-relativistic ($\Gamma = 4/3$) and Newtonian
($\Gamma = 5/3$) regimes.  The reader is referred to \citet{paperI}
for a detailed description of the code and an extensive list of
numerical tests.

For the initial conditions we use a conical wedge of half-opening
angle $\theta_0$, within which the initial radial profiles of
pressure, density and Lorentz factor in the post-shock region are
taken from the spherical  Blandford-McKee self-similar solutions for a stratified
medium: 
\begin{equation}
 \rho = A_k r^{-k} \;.
\end{equation}
Two-dimensional simulations with $k=0$
(homogeneous medium), $k=1$ and $k=2$
(corresponding to a steady stellar wind medium) are then evolved to
study the lateral expansion and deceleration of the jet.

To accurately study the dynamics near the jet break time, an initial
shock Lorentz factor of $\Gamma_{\rm sh,0} =\sqrt{2}\times 20$ and an
initial half-opening jet angle $\theta_0 = 0.2$~rad are selected, so
that $\Gamma_{\rm sh,0}\gg\theta_0^{-1}$. The isotropic equivalent
energy is taken to be $E_{\rm iso} = 10^{53}$~erg, corresponding to a
total jet energy content  of $E_{\rm jet} = E_{\rm iso} (1-\cos \theta_0) \sim 2
\times 10^{51}$~erg.  The ambient medium is assumed to have  a density $\rho_0 = A_0 =
1.67 \times 10^{-24}$~g cm$^{-3}$ (for the case $k=0$, which
corresponds to $\rho_0 = n_0m_pc^2$ with $n_0 = 1\;{\rm cm^{-3}}$),
and a pressure $p = \eta \rho_0 c^2$, with $\eta=10^{-10}$. The value
of $\eta$ has no bearing on the outcome of the simulation as long as
the Mach number remains large, i.e.  $\mathcal{M} \sim \eta^{-1/2}
v_{\rm sh}/c \gg 1$, where $v_{\rm sh}$ is the shock velocity.
As the simulation continues to evolve well into
the Newtonian regime, this condition can be expressed as $v_{\rm sh}
\gg 3\; (\eta/10^{-10})^{1/2}\;{\rm km\;s^{-1}}$.
The density profiles in the cases $k=1,2$ are fixed here by 
assuming the jet break radius (in the lab frame) to be the same for all
$k$: $R_{\rm j}(k)=R_{\rm j}(k=0)$. This can be rewritten \citep{BM76} as
\begin{equation}
R_{\rm j} = \left( \frac{(17-4k)E_{\rm iso}}{8\pi A_k \Gamma_{\rm j}^2 c^2} \right)^{1/(3-k)} 
          = \left( \frac{17 E_{\rm iso}}{8\pi A_0 \Gamma_{\rm j}^2 c^2} \right)^{1/3}\ ,
\label{eq:rjb}
\end{equation}
where $\Gamma_{\rm j}=\sqrt{2}/\theta_0$.

From equation (\ref{eq:rjb}) we have $ A_k = A_0
R^k_{\rm j} (17 - 4 k)/17 $, so that the density of the ambient medium
is given by
\begin{equation}
  \rho = \frac{17 - 4 k}{17}A_0 \left(\frac{r}{R_{\rm j}}\right)^{-k}  \;,
  \label{eq:dec0}
\end{equation}
which guarantees  $R_{\rm j}$ to remain unchanged for varying $k$. 
With this constraint, the value of the density at the jet break radius
$\rho( r = R_{\rm j})$ differs, compared to the $k=0$ case, by factors of $13/17$ and
$9/17$ for $k=1$ and $k=2$, respectively.
In the simulations presented in this paper, $R_{\rm j} = 9.655 \times 10^{17}$~cm,
corresponding to a jet break time of $t_j = R_{\rm j}/c \approx 372$~days.\\

The jet is expected to begin decelerating to non-relativistic speeds at
\begin{equation}
  t_{\rm NR} \approx \frac{L_{\rm Sedov}}{c} = 
    \left( \frac{(3-k)E_{\rm iso}}{4 \pi A_k c^2} \right)^{1/(3-k)}  \;,
  \label{eq:sed}
\end{equation}
corresponding to $t_{\rm NR} \approx 970$, $3800$ and $11000$~days (in the lab frame) 
for $k=0$, 1, 2 respectively.  

The simulations with $k=0,1$ employs a spherical computational domain of radial and angular 
size $(L_r, L_\theta) = (1.1\times 10^{19}$~cm, $\pi/2)$ while the simulation
with $k=2$ uses $(L_r, L_\theta) = (2.2\times 10^{19}$~cm, $\pi/2)$. The inner 
boundaries are located at $(1.8,1.2,0.3) \times 10^{17}$~cm for $k=(0,1,2)$, respectively.  
The AMR code uses a basic grid of $(100,6)$ cells in
the $(r,\theta)$ directions, and 15 ($k=0,1$) or 16 ($k=2$) levels of 
refinement, corresponding to a maximum resolution of $(\Delta r_{\rm min},
\Delta\theta_{\rm min})=(6.71 \times 10^{12}$~cm, $1.60\times 10^{-5}$~rad).
To keep  the resolution of the 
relativistic thin shell $\Delta \propto t^{4-k}$ approximately constant, the maximum 
number of levels of refinement $N_{\rm levels}$ is decreased with 
time \citep{paperI} as $N_{\rm levels} = \max[7, N_{\rm levels, 0} -
(4-k) \log(t/t_0)/\log(2)]$. 
The simulations are halted  after $150$~years.
We also carried out a higher resolution simulation (for the $k=2$ case) 
using a basic grid of $(1000,16)$ cells in the $(r,\theta)$ directions, 
and 14 levels of refinement. 
The light curves computed from this simulation are very similar to those  
those obtained from the lower resolution run, implying that convergence 
has been achieved.

The \emph{Mezcal} code is parallelised using the ``Message Passing 
Interface''  (MPI) library, enabling the highest resolution simulation to 
be run in about two weeks on a local supercomputer with 160 processors,
and the low resolution in about a quarter of that time.

\subsection{Afterglow Radiation} 

To compute the afterglow radiation, we use the method described in
\citet{paperI}. As the main goal of the current calculations 
is to study the effect of the jet dynamics on the afterglow
lightcurves, a simple model is employed to calculate the emanating
radiation.  It assumes synchrotron to be the primary emitting
mechanism, while ignoring self-absorption and inverse Compton
scattering. Furthermore, a simple prescription for electron cooling
\citep{paperI} is assumed, which is similar to the one used by
\citet{Granot01} and \citet{ZM09}.

In addition to the contributions to the afterglow radiation computed
by post-processing the results of the hydrodynamics simulations, 
contributions from earlier lab frame times are included,
corresponding to the blast-wave decelerating from $\Gamma_1 =
\Gamma(\chi = 1) =\Gamma_{\rm sh}/\sqrt{2} = 200$ to $\Gamma_1 = 20$. Here
$\chi(r/R_{\rm sh}) = 1 + 2 (4-k) \Gamma_{\rm sh}^2\left(1-r/R_{\rm sh}\right)$
is a self-similar variable which quantifies the distance from the shock 
front \citep{BM76}.
These are computed using the same conical wedge taken out of the  Blandford-McKee
self-similar solution that is used for initializing our
simulations. The mapping of the
 Blandford-McKee solution is implemented by using a high resolution grid, starting at
the position of the shock front (which varies with time) and sampling
the  Blandford-McKee solution at intervals of fixed $\Delta \Gamma = 0.01$. 
The values of the proper density $\rho$, internal energy density 
$e_{\rm int}$, 4-velocity $u$ and self-similar variable $\chi$ replace 
those coming from the simulations, and are taken from the 
Blandford-McKee self-similar solution at the relevant lab frame time. 
In order to calculate the contributions to the
observed radiation, the {\it mapped} jet radial structure is subsequently
integrated over all angles  ($0\leq\theta\leq \theta_0$; $0\leq\phi\leq 2\pi$).
This procedure provides a reasonable description of the afterglow 
radiation at earlier times and it is significantly more accurate than 
ignoring the contributions from lab frame times preceding the start of 
the simulation.

The microphysics processes responsible for field amplification and
particle acceleration are parametrized here by assuming that the
magnetic field everywhere in the shocked region holds a fraction
$\epsilon_B = 0.1$ of the local internal energy density in the flow,
while the non-thermal electrons just behind the shock hold a
fraction $\epsilon_e = 0.1$ of the internal energy, and have a
power-law energy distribution, $N(\gamma_e)\propto\gamma_e^{-p}$, with
$p=2.5$. We also assume the source to be at a redshift of
$z=1$, corresponding to a luminosity distance of $d_L = 2.05\times
10^{28}\;$cm.  The afterglow radiation code has been tested in
\citet{paperI}. The simulation with $k=0$, in particular,
gives afterglow light curves that are nearly identical to those computed
by \citet{ZM09}.

\section{Jet Dynamics in a Stratified Medium}\label{dynamics}

\begin{figure}
 \includegraphics[width=0.5\textwidth]{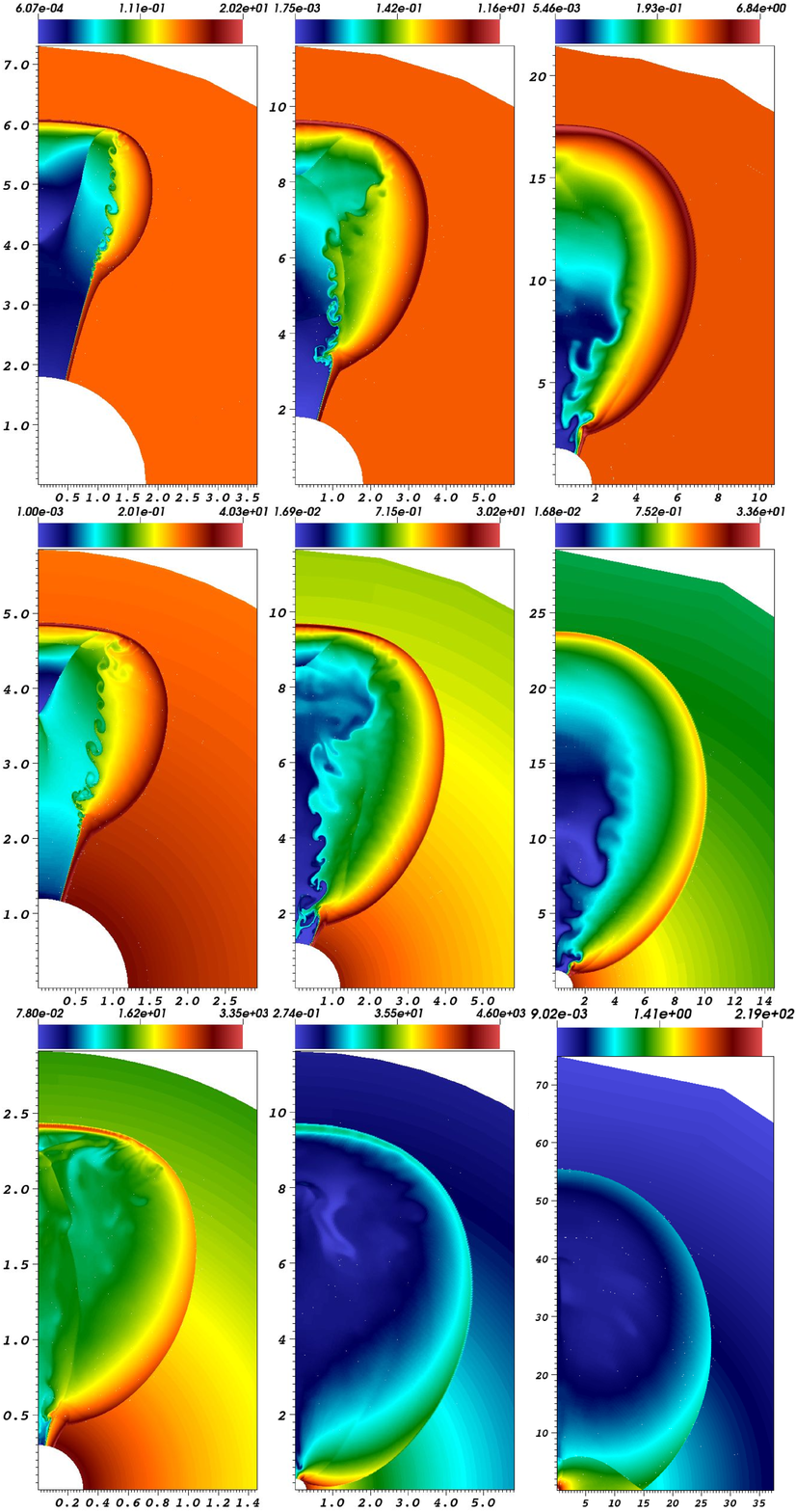}
\caption{The temporal evolution of GRB jets in a stratified medium 
with $k=0,1,2$ (top to bottom panels, respectively).  The three
plotted times, whose exact values dependent on $k$,
have been selected so that the  Blandford-McKee Lorentz factor in the post-shock region
$\Gamma(\chi=1)$ is equal to 10, 5 and 2 (left to right).
 Shown are logarithmic lab frame density cuts in cm$^{-3}$. 
Calculations were done in two-dimensional spherical
coordinates with the axes corresponding to the $r-$ and $z-$
directions in units of 10$^{17}\;$cm. The position of the shock
front corresponding to a $\Gamma(\chi=1) = 5$ is the same for all $k$ values,
consistently with the normalization used in the simulations
(see equation \ref{eq:rjb}).}
\label{fig1}
\end{figure}

\begin{figure}
 \includegraphics[width=0.5\textwidth]{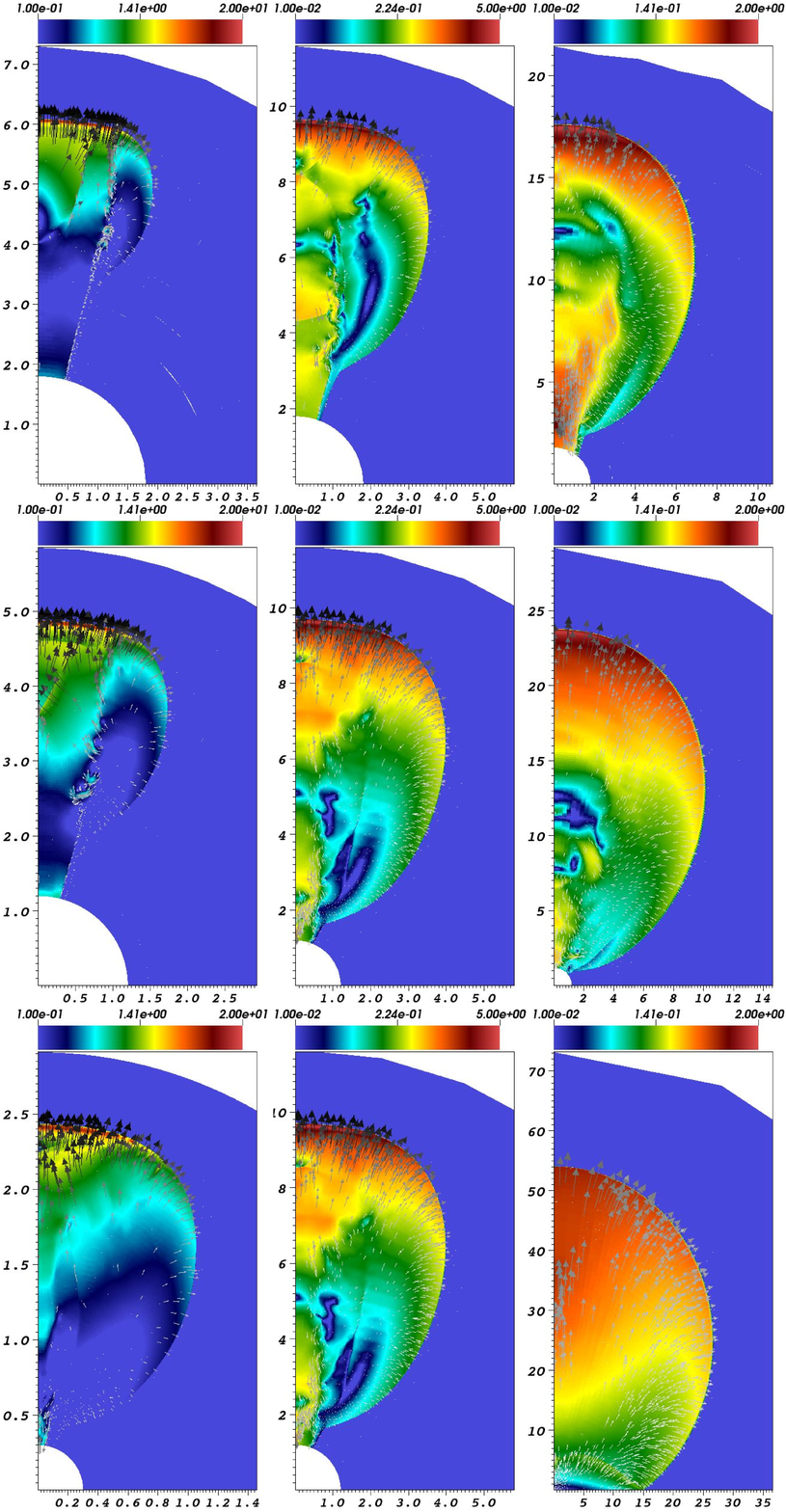}
\caption{The same  evolutionary sequence  depicted in  
Figure~\ref{fig1} but for the absolute value of the velocity
quadrivector. The superposed velocity field arrows are represented by
a {\it gray scale} color scheme linear with respect to the 3-velocity,
with dark corresponding to speeds $\sim c$ and lighter to $v \ll c$.}
\label{fig2}
\end{figure}

Detailed hydrodynamic simulations of the evolution of a GRB jet in a
stratified medium with $k=0,1,2$ are presented in Figures~\ref{fig1}
and \ref{fig2} where the density and velocity contours of the
expanding ejecta at various times are plotted.  A transient phase
caused by the sharp lateral discontinuity in the initial conditions is
observed in all cases as the shock expands laterally and a rarefaction
front moves towards the jet axis. This initial phase, during which
shearing instabilities are observed to be prominent at the contact
discontinuity (separating the {\it original}  Blandford-McKee wedge material and the
later shocked external medium), lasts for about a dynamical timescale
and is followed by the establishment of an egg-like bow shock
structure that persists throughout the simulations.
The velocity quadrivector (Figure~\ref{fig2}) shows strong 
stratification in the $\theta$ direction. The expansion velocity of the 
jet remains mainly radial at most angles, with a non-relativistic 
angular component being prominent at large angles.
The  substructures seen in the velocity quadrivector  along the $z-$axis 
at late times (generated 
by the convergence of turbulent flow)   carry a small fraction of the energy and have a 
negligible effect on the lightcurves.

\begin{figure}
 \includegraphics[width=0.25\textwidth,angle=90]{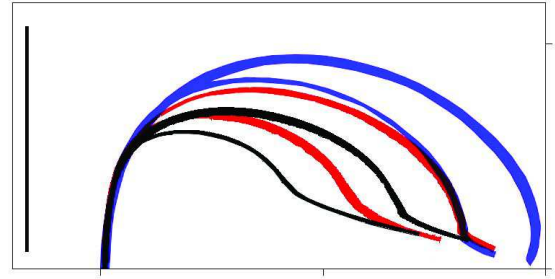}
\caption{A comparison between the bow shock structures  
depicted in Figure \ref{fig1} for $k=0$ (black), $k=1$ (red) and $k=2$
(blue). The two times have been selected so that the jet has the same
Lorentz factor of 10 and 5 in all simulations.  The evolutionary scale
unit of ${1 \over 2} ct$ is indicated with a black transverse bar.
The origin of the axis is located at the right bottom corner and
the jets main direction of propagation is toward negative $x$ in this
figure. The simulations are normalized with respect to $ct$.}
\label{fig3}
\end{figure}

Similar resulting bow shock structures are observed for $k=0$, $1$ and
$2$. However, because the rate at which mass is swept-up is larger for 
smaller values of $k$, the bow shock lateral expansion augments with
increasing $k$.  As clearly seen in Figure~\ref{fig3}, the ratio
between the bow shock width and height as the ejecta expand changes
with $k$. This can be understood as follows. Small values of
$k$ correspond to a larger increase in the swept-up external mass and 
larger decrease in the Lorenz factor. For the spherical 
case, in particular, $M(<R) \propto R^{3-k}$ and $\Gamma\propto R^{-(3-k)/2}$, 
and the same trend should persist for the non-spherical case.

\begin{figure}
 \includegraphics[width=0.45\textwidth]{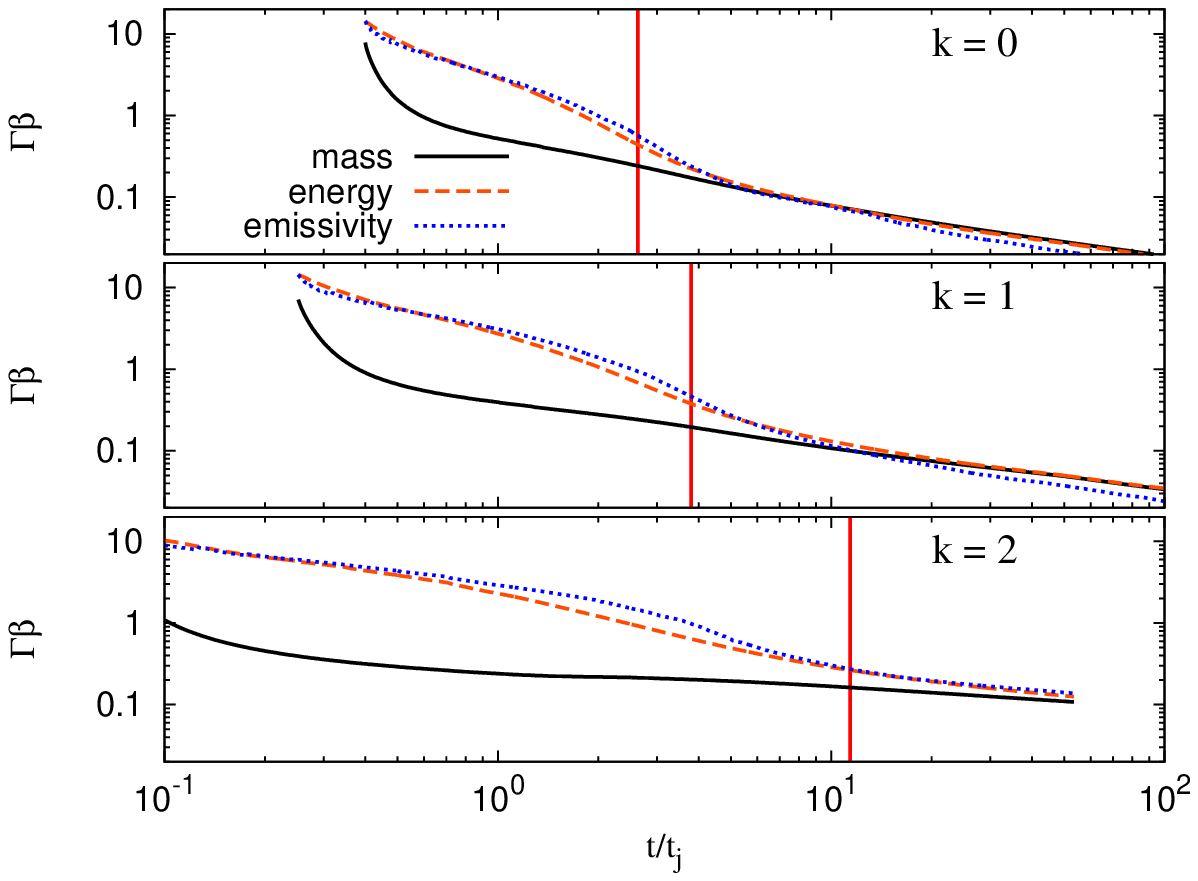}\\
 \includegraphics[width=0.45\textwidth]{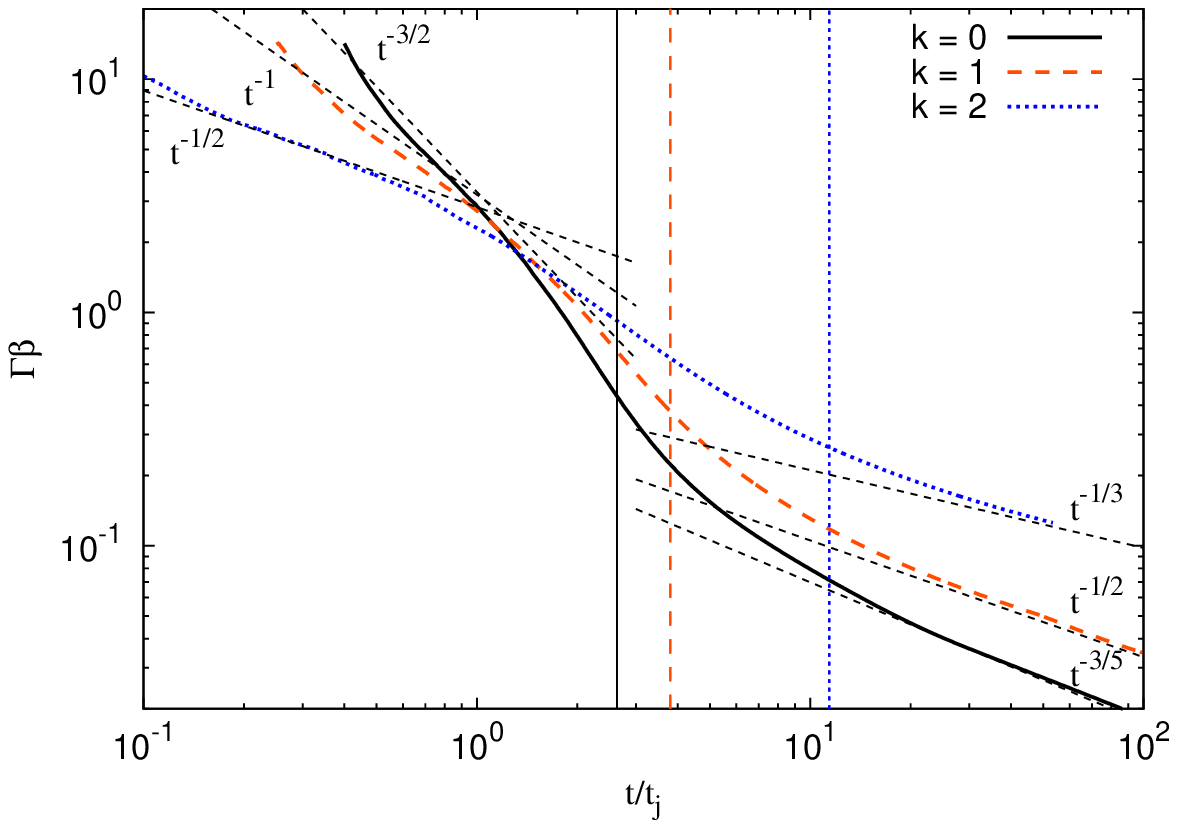}
\caption{The  temporal evolution (in the lab frame) of the velocity 
quadrivector, $u = \Gamma v/c = \Gamma\beta$, in units of the jet
break time.  {\it Upper panel}: The different lines give the evolution
of $u$ within the jet when averaged over (rest-) mass, energy
(excluding rest mass) and over the emissivity (or contribution to the
observed flux for a distant observer along the jet axis) at
10$^{17}\;$Hz for the evolutionary sequences shown in
Figure~\ref{fig1}.  The non-relativistic transition time, $t_{\rm NR}(E_{\rm
iso})$, is shown in the figure as solid vertical lines for
$k=0,1,2$.
\emph{Lower panel}: A comparison between the velocity quadrivector $u(t)$
averaged over energy for $k=0,1,2$.  The non-relativistic transition time, $t_{\rm
NR}(E_{\rm iso})$, is shown as a thin vertical line with the same
line-style and color as the thick lines for the corresponding $u(t)$.
The Blandford-McKee and Sedov-Taylor self-similar solutions are
plotted as black thin dashed lines together with the corresponding
$-d\log u/d\log t$ slopes.  
}
\label{fig4}
\end{figure}

The velocity quadrivector, $u = \Gamma v/c = \Gamma\beta$, of the
expanding jets are shown in Figure \ref{fig4} for three different
angle-integrated quantities: mass, energy and emissivity. 
The mean value of $u$ is larger when weighted over the energy or
emissivity than over the shocked rest-mass until $t \lesssim 10 \times t_j$. 
This clearly illustrates, in agreement with previous analytical and numerical
results limited to the case $k=0$ \citep{Granot01,ZM09}, that during the
relativistic phase, most of the shocked 
rest mass resides in relatively slow material at the edges of the jet, while 
most of the energy is stored in the fastest moving material near the head of the jet.

As illustrated in Figure~\ref{fig4}, the Blandford-McKee and Sedov-Taylor self-similar
solutions fail to provide an adequate description of the jet dynamics
at $t_j \lesssim t \lesssim t_{\rm NR}(E_{\rm iso})$ with the
disagreement becoming less pronounced before $t_j$ and after $t_{\rm
NR}(E_{\rm iso})$. Between these two limiting cases, $-d\log u/d\log
t$ evolves at early and late times between the two asymptotic slope values, 
as seen in the bottom panel of Figure~\ref{fig4}.
The evolution of $-d\log u/d\log t$ is, however, non-monotonic as it  first increases
above $(3-k)/2$  and  only then decreases down to $(3-k)/(5-k)$. This behavior 
is mainly caused by the faster decrease in $\Gamma$ compared to a spherical
flow at $t>t_j$ due to  the lateral expansion of the jet. It also relates to the fact that the Blandford-McKee solution {depends on} $E_{\rm
iso}$ while the corresponding Sedov-Taylor solution uses   the jet's true energy,
$E_{\rm jet}$ and, as a result, the ratio of $u(t)$ for these two  limiting cases is $\sim \theta_0^{-1}$ at $t = t_{\rm
NR}(E_{\rm jet})$ and $\sim\theta_0^{-2/(5-k)}$ at $t = t_{\rm
NR}(E_{\rm iso})
\sim \theta_0^{-2/(3-k)}t_{\rm NR}(E_{\rm jet})$.

\begin{figure}
 \includegraphics[width=0.45\textwidth]{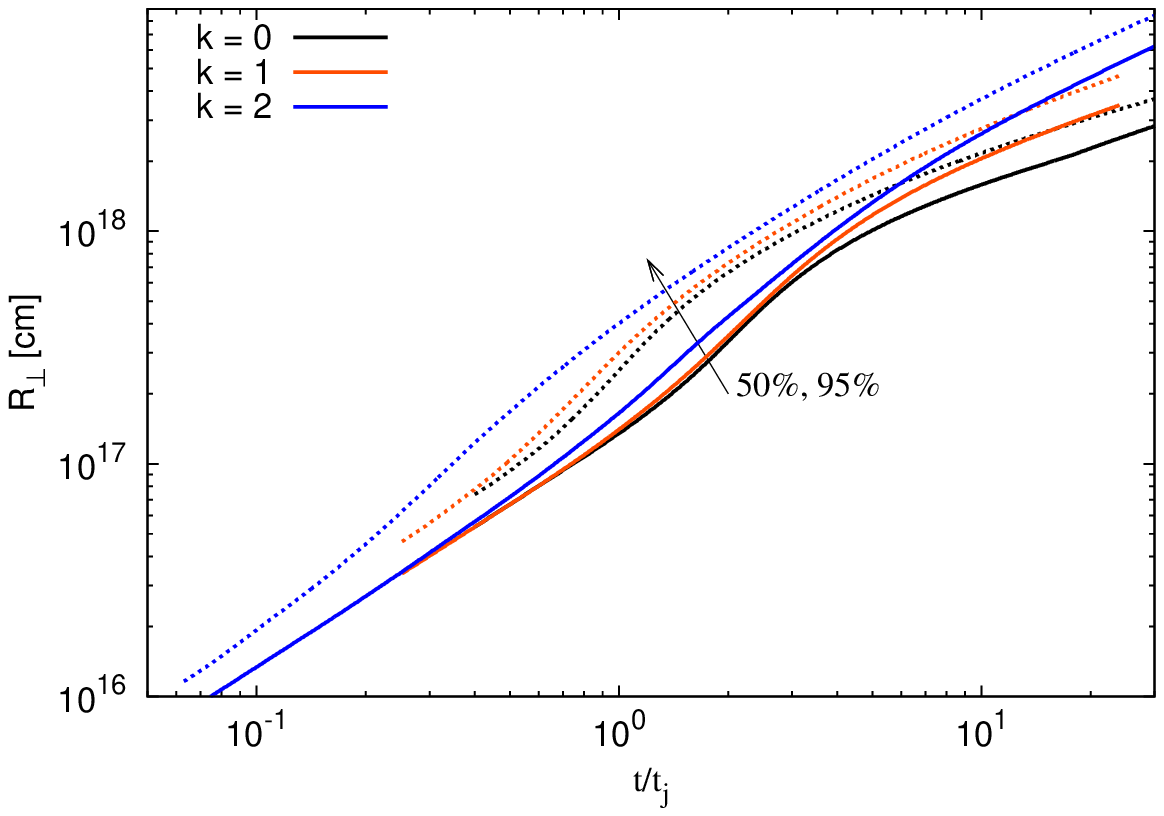}\\
 \includegraphics[width=0.45\textwidth]{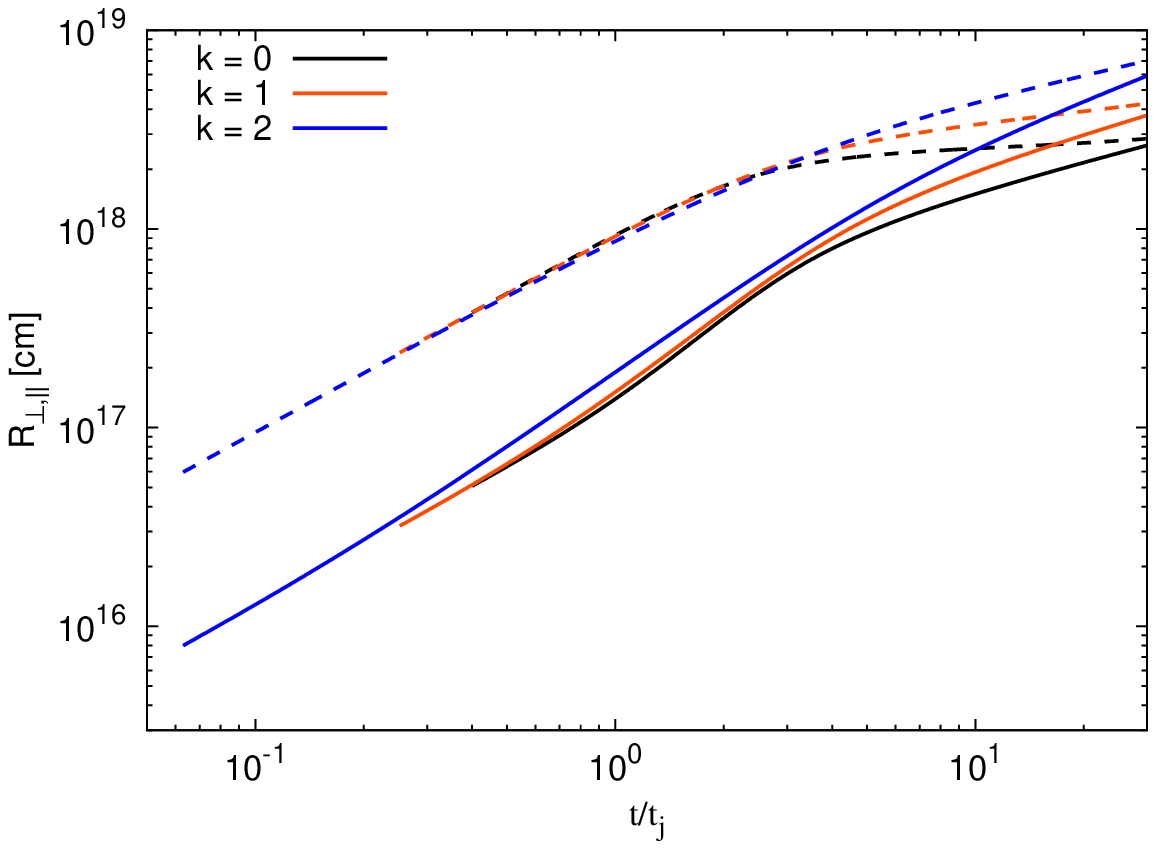}\\
  \includegraphics[width=0.45\textwidth]{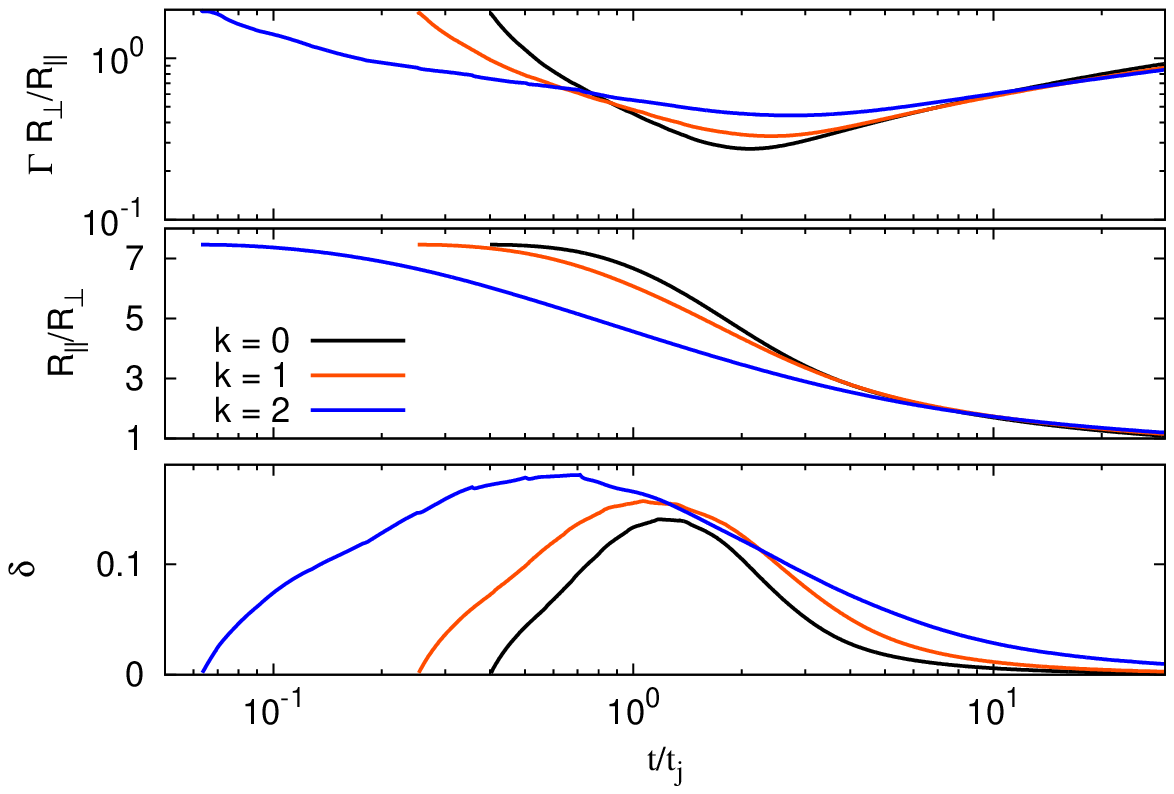}
\caption{Temporal evolution (in the lab frame) of  $R_\perp$,  
$R_\parallel$, $(\Gamma R_\perp/ R_\parallel)$,
$(R_\parallel/R_\perp)$ and $\delta=d(R_\perp/R_\parallel)/dt$
in units of the jet break time. Here $R_\perp$ and
$R_\parallel$ are the transverse (cylindrical radius) and parallel
(along the $z$ axis) scales of the expanding jet, respectively.
The different lines give the evolution (top panel) of $R_\perp$ 
defined as the transverse scale of the jet that contains
50\% (solid) or 95\% (dashed) of the total energy excluding rest
mass (top panel) and the evolution (middle, bottom panels) of 
$R_\perp$ ($R_\parallel$) averaged over the total energy excluding 
rest mass.  
}
\label{fig5}
\end{figure}

\begin{figure}
 \includegraphics[width=0.45\textwidth]{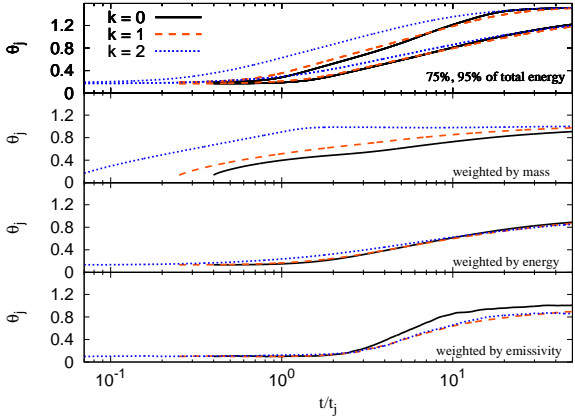}
\caption{ The evolution of the jet half-opening angle $\theta_j$ as 
a function of time for different $k$ values. The various panels give
$\theta_j$ derived based on the angle-integrated mass, energy
(excluding rest mass) and emissivity (at 10$^{17}$~Hz).  Also plotted
is the evolution of $\theta_j$ computed as the characteristic angular
scale containing 75\% or 95\% of the total energy (excluding rest
mass).}
\label{fig6}
\end{figure}

Figures \ref{fig5} and \ref{fig6} show the resulting $R_\perp(t)$,
$R_\parallel(t)$ and $\theta_j(t)$ for $k=0,1,2$ and different recipes
for estimating the transverse, parallel and angular size scales within
the jet (e.g. when averaged over mass, energy and emissivity).  
For all values of $k$ the early lateral spreading of the jet,
which starts around $t \sim t_j$, is observed to initially involve
only a modest fraction of the total energy, with the bulk of the
energy reaching angles well above $\theta_0$ at significantly later
times.

For $k=0$, previous numerical simulations and analytical models
assuming a small lateral expansion for $t\sim t_{\rm NR}$ 
\citep[e.g.][]{GR-RL05} have shown that spherical symmetry is 
approached on timescales much larger than $t_{\rm NR}$.
In particular, Figures \ref{fig5} shows that the growth of $R_\parallel$ is 
essentially stalled at $t\sim t_{\rm NR}$ while $R_\perp$ continues to 
grow as the flow gradually approaches spherical symmetry.
For increasing $k$ this effect is less pronounced,
since $R_\parallel$ continues to increase even after 
$t_{\rm NR} (E_{\rm iso})$, albeit more slowly. 
This contributes to the faster growth in $\theta_j$ for lower $k$-values 
at late times, contrary to the opposite situation at early times ($t\lesssim t_j$).
This causes GRB jets expanding into steeper density profiles to
approach spherical symmetry at progressively later times as argued by
\citet{2010ApJ...716.1028R} for $k=2$.

Since the rate of lateral spreading of the jet increases as  
$\Gamma$ decreases (see, e.g., equation 2 of \citealt{Granot07}) and
$\Gamma(R_j)=\theta_0^{-1}$ is the same for all $k$, then the jet
lateral spreading is expected to increase with $k$ for $R\lesssim R_j$
(where $\Gamma(R)$ decreases with $k$ for a given $R$), while the
opposite should hold for $R\gtrsim R_j$ (where, for a given $R$,
$\Gamma(R)$ increases with $k$). Such a behavior is also seen in analytic
models \citep{Granot07,GP11}.

Figure \ref{fig5} also plots the temporal evolution of $\Gamma
R_\perp/R_\parallel \approx \Gamma\theta_j$, which is observed to
approach unity at $t\gg t_j$. This should be compared with the
results of semi-analytic models \citep[e.g.]{1997ApJ...487L...1R,SPH99,KP00}. These models predict
$\Gamma\theta_j \approx 1$ at $t\gtrsim t_j$, and $\Gamma$ to decrease 
rapidly with lab frame time $t$, which is not
observed here.  In the simulations, $\Gamma$ decreases rather slowly
with $t$ (as a power-law). The jet angular size $\theta_j$ (see Figure 
\ref{fig6}), on the other hand, is observed to increase only logarithmically 
with $t$ for all $k$ until the flow becomes non-relativistic. 

As shown in Figure \ref{fig6}, the weighted mean of $\theta_j$ 
over the emissivity (and to a slightly lesser extent over the energy) 
remains practically constant until $t/t_j \sim$ a few, while the 
weighted mean over the shocked rest mass is significantly larger, 
in accord with earlier results \citep{Granot01, 2001grba.conf..300P,ZM09}.  
This indicates that, as argued before, a large fraction of the 
swept-up external rest mass is concentrated at the edges of the jet, 
while most of the energy and emission lies near the head. 
Moreover, it implies that (as discussed above and seen in the temporal evolution of   
$\delta$ depicted at the bottom of Figure~\ref{fig5}) the lateral expansion 
at early times, $t\lesssim t_j$, is significantly faster for larger 
values of $k$, while the situation is reversed at late times. 

\begin{figure}
 \includegraphics[width=0.4\textwidth]{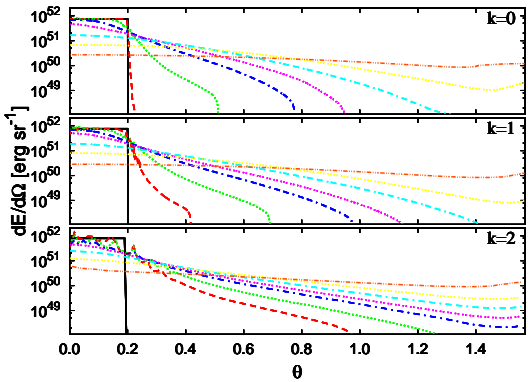}
\caption{ The angular distribution of the energy content in the 
expanding jet (excluding rest mass) at $t=0$ (black line), 0.5 (red), 
1 (green),  2 (dark blue),  5 (pink), 10 (light blue), 20 (yellow),
50 (orange) ~years. }
\label{fig7}
\end{figure}

Figure~\ref{fig7} plots the temporal evolution of the energy
(excluding rest energy) per solid angle, $\epsilon = dE/d\Omega$, as a
function of the angle $\theta$ from the jet symmetry axis, for
$k=0,1,2$. At $t \gtrsim 50$~yrs the energy distribution 
appears nearly spherical for all $k$s. At earlier times, a
 clear $k$-dependence trend is observed, where 
the energy spreads to larger solid angles faster for a more 
stratified medium, but a correlation is
less evident when one compares $\epsilon(\theta)$ for different
$k$-values at the same  four velocity $u$ rather than the
same lab frame time $t$.

Abundant confirmation is provided here that the dynamics of
GRB jets are greatly modified by the radial profile of the
surrounding circumburst density.  Most analytic formalisms
\citep[e.g.,][]{Rhoads99}  derive an exponential lateral
spreading with lab frame time or radius at $t > t_j$, which ultimately
erases all information about the initial jet opening angle and relies
solely on the true energy content of the jet: $E_{\rm jet}$. No
exponential lateral expansion is observed in our study for $k =
1,\,2$, consistent with previous numerical work for expansion in a
constant density medium ~\citep{Granot01,ZM09,vEM11}. As illustrated
in Figure~\ref{fig6}, the evolution of the jet's angular scale
containing a constant fraction of the total energy is logarithmic and
is not self-similar as it retains memory of the initial jet opening
angle. 
The deviation from the expected self-similar exponential lateral expansion 
behavior ~\citep{Gruzinov07} might be at least partly due to $u$ rapidly
decreasing with the polar angle $\theta$ from the jet symmetry axis,
so that the flow is no longer ultra-relativistic ($u\gg 1$) as
it has been previously assumed. Even with the expectation that such a self-similar
solution would be only very slowly attained
\citep{Gruzinov07}, the maximal Lorentz factor at the head of the jet
in this formalism is predicted to decrease exponentially with time,
which appears to be inconsistent with our numerical results. 

The resolution of this apparent inconsistency between analytic
models and numerical simulations can be attributed to the modest
values of $\theta_0$ used in the simulations, which result in the
breakdown of the analytic models, which assume $\Gamma\gg 1$ and
$\theta_j\ll 1$ soon after the jet starts spreading sideways
$\Gamma<\theta_0^{-1}$) and before it can reach a phase
of exponential lateral expansion \citep{WWF11,GP11}. In the small 
region in which the analytical models are valid:$1\ll\Gamma<\theta_0^{-1}$, 
there is reasonable agreement with   simulation results \citep{WWF11}. 
A generalization of these analytic models to any values of
$\Gamma$ or $\theta_j$ \citep{GP11}  shows reasonable agreement  
with  the results of  simulations  from the early ultra-relativistic stage
to the late Newtonian stage. Such generalized analytic
models predict that if the jet is initially extremely narrow then there
should still be an early phase of exponential lateral
spreading. However, these models make the simplifying approximation 
of a uniform jet, while in practice $u$ quickly drops with $\theta$. 
This causes a breakdown of the $u\gg 1$ assumption used to derive the 
self-similar solution, which is only slowly attained even under ideal 
conditions \citep{Gruzinov07}.


\section{Afterglow Lightcurves}\label{rad}

\begin{figure}
 \includegraphics[width=0.4\textwidth]{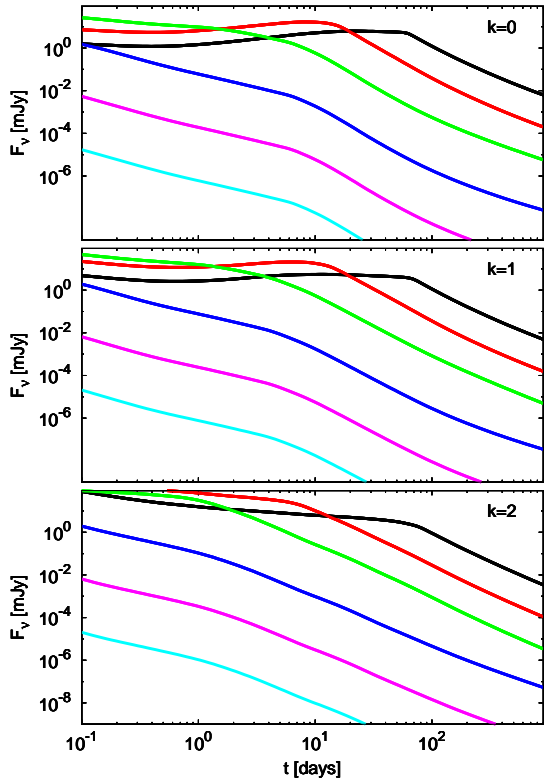}
 \includegraphics[width=0.4\textwidth]{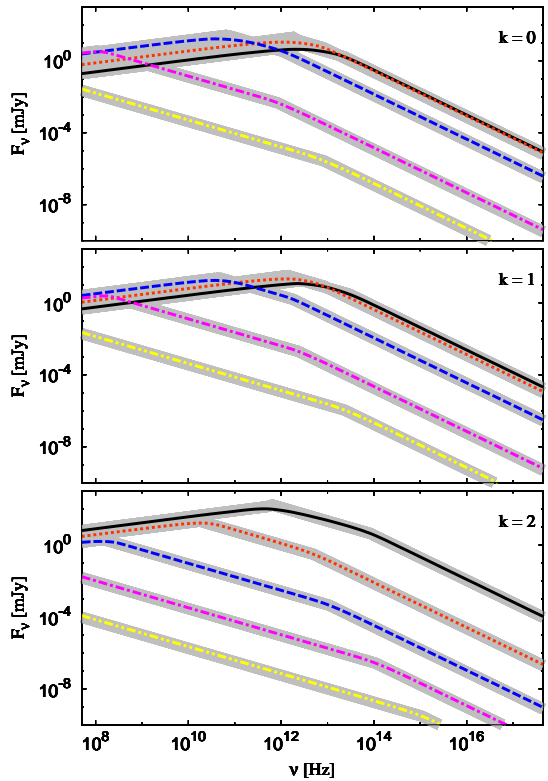}
\caption{
Light curves at $\nu = 10^9$, $10^{11}$, $10^{13}$, $10^{15}$,
$10^{17}$, $10^{19}\;$Hz (black, red, green, blue, purple, cyan
respectively; {\it top panels}) and spectra at $t_{\rm
obs}=0.1,\,1,\,10,\,100,\,1000$~days (black/continuous line, 
orange/dotted, blue/dashed, purple/dash-dotted, yellow/dash-dash-dotted;
{\it bottom panels}) for the models $k=0,1,2$ (top to bottom panels),
calculated including electron cooling and the contribution from a
mapped Blandford-McKee solution (with $20\leq\max(\Gamma)\leq
200$).}
\label{fig8}
\end{figure}

\begin{figure}
 \includegraphics[width=0.5\textwidth]{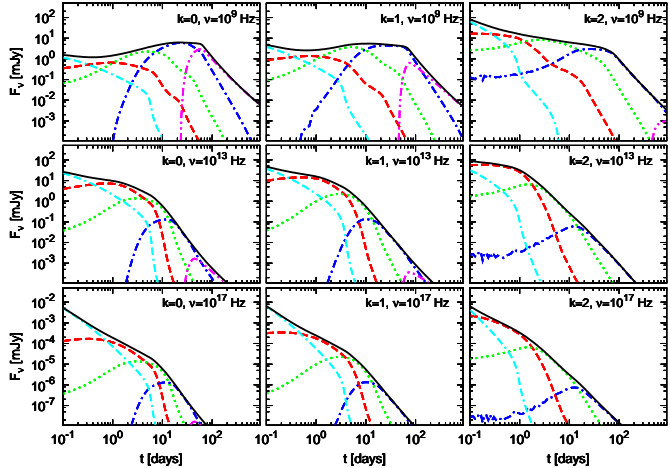}
\caption{
   Afterglow lightcurves emanating  at different Lorentz factors. The red, green, blue, purple
   (dashed, dotted, dashed-dotted, dashed-dotted-dotted) lines
   are the contributions to the total light curve (in black) 
   computed by using the outputs of the simulations 
   at the lab frame times where $\Gamma_{\rm sh}(t)/\sqrt{2}$ 
   (as given by the  Blandford-McKee solution) ranges between 10 and 20, 5 and 10,
   2 and 5, 1 and 2 respectively. The cyan dashed-dotted lines are the
   contributions from a  Blandford-McKee wedge with $20 \leq \Gamma_{\rm sh}(t)/\sqrt{2}
   \leq 200$. The light curves include electron cooling.}
\label{fig9}
\end{figure}

Figure~\ref{fig8} shows the emerging  light curves at frequencies ranging from the
radio to gamma-rays ($\nu = 10^9$, $10^{11}$, $10^{13}$, $10^{15}$,
$10^{17}$, $10^{19}\;$Hz), and  corresponding  spectra at different observed times $t_{\rm
obs}$, for $k = 0,\,1,\,2$, including the effects of electron cooling
and the contribution from a mapped Blandford-McKee solution (with
$20\leq\max(\Gamma)\leq 200$). Figure~\ref{fig9} shows the light
curves computed for $\nu = 10^9$, $10^{13}$ and $10^{17}\;$Hz, for the
two dimensional simulation and the  Blandford-McKee conical wedge, as in
Figure~\ref{fig8}, but illustrating the contributions to the
lightcurve arising from the various evolutionary  stages of the blast
wave, quantified here by considering the emission from lab frame
times where $\Gamma_{\rm sh}(t)/\sqrt{2}$, given by the Blandford-McKee 
solution, ranges between 10 and 20, 5 and 10, 2 and 5, 1 and 2 
respectively. As expected, lower Lorentz factors
contribute to the observed flux at later times. A slightly more subtle
effect is that at the same observed time the flux at low frequencies
comes from slightly later lab frame times $t$ (corresponding to a
lower  Blandford-McKee Lorentz factor $\Gamma_{\rm sh}(t)$). This is because 
there is a lower flux contribution from the sides of the jet compared to
the center, as reflected by the fact that the afterglow image is
more limb-brightened at higher frequencies and less so at lower
frequencies \citep{GPS99, 2001ApJ...551L..63G,Granot08}, resulting
in a smaller typical angular delay time ($t_\theta = R(t) \approx
R\theta^2/2c$) in the arrival of photons to the observer (which is
along the jet axis in these figures). As a result, the flux at the 
same observed time $t_{\rm obs}$ is dominated by larger lab frame times $t$.

The spectra at different observer times are shown in Figure \ref{fig8}.
For all values of $k$, the spectra evolves from a fast cooling
(with $\nu_c<\nu_m$ and $F_\nu \propto \nu^{1/3}$,
$\nu^{-1/2}$, $\nu^{-p/2}$ for $\nu < \nu_c$,  $\nu_c < \nu < \nu_m$,
 $\nu > \nu_c$ respectively) to a slow cooling regime (with 
$\nu_m<\nu_c$ and $F_\nu \propto \nu^{1/3}$,
$\nu^{(1-p)/2}$, $\nu^{-p/2}$ for $\nu < \nu_m$,  $\nu_m < \nu < \nu_c$,
 $\nu > \nu_m$ respectively). 
The characteristic frequency $\nu_m$ quickly drops with time with an 
asymptotic slope of $-2.9$, $-2.6$, $-2$ for $k=0,1,2$ respectively,
(while one expects $\nu_m \propto t^{-(15-4k)/(5-k)}$, which is 
relatively closed to our result),
while $\nu_c$ increases at late times as $\nu_c \propto t$, that is, 
with a slope independent on the particular stratification
of the ambient medium (for comparison, in the Sedov-Taylor regime 
one expects $\nu_c \propto t^{(2k-1)/(5-k)}$). 
%

As shown in \S \ref{dynamics}, a jet moving in a stratified 
medium (with $k=1$ and $k=2$) decelerates to sub-relativistic speed 
over larger distances with respect to a jet moving in an homogeneous 
medium ($k=0$). The consequences of it on the light curve are particularly
evident at radio frequencies (Figure~\ref{fig9}), where the contribution
from mildly- and sub-relativistic material is negligible in the $k=2$
case and dominant in the $k=0$ up to $t \sim 10^3$~days.

\begin{figure}
 \includegraphics[width=0.4\textwidth]{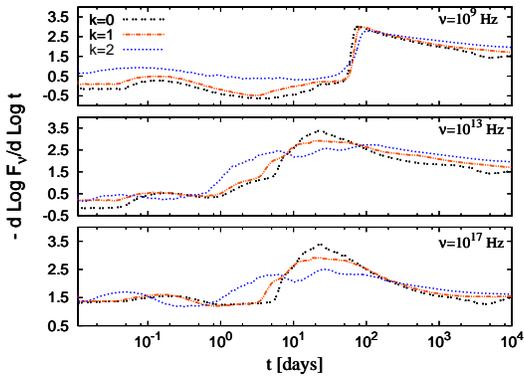}
\caption{
``Shape of the jet break'', i.e. temporal decay of light curve, given
by $\alpha\equiv- d\log F_\nu/d\log t_{\rm obs}$ as a function of
$t_{\rm obs}$, at three different frequencies, including 
electron cooling.}
\label{fig10}
\end{figure}

Figures~\ref{fig8} and \ref{fig9} show a pan-chromatic dip or
flattening in the lightcurves at around half a day for $k = 0$, a
third of a day for $k = 1$ and significantly earlier for $k = 2$. This
feature is also seen in Figure~\ref{fig10}, which shows the temporal
index $\alpha\equiv- d\log F_\nu/d\log t_{\rm obs}$ as a function of
$t_{\rm obs}$, where the earliest value of
$\alpha$  is larger than expected analytically
for a spherical flow (or for a jet viewed along its axis, before the jet
break time). Figure~\ref{fig9} clearly illustrates the reason for this behavior. It
basically occurs at the point where the dominant contribution to the
observed flux switches from the  Blandford-McKee wedge with $20 \leq
\Gamma_{\rm sh}(t) \leq 200$ to the simulation, which corresponds to
later lab-frame times. As pointed out and calculated  in   \citealt{paperI} for
the spherical case, the relaxation of the mapping of the analytic  Blandford-McKee
self-similar solution to the numerical solution and the finite
resolution of the simulation result in a dip in the Lorentz factor
that is gradually recovered as the shocked region becomes wider and
thus better resolved with time. This produces a dip in the
lightcurve, that gradually goes away as the resolution of the
simulation is increased (see Figures 5, 6, 7 of  \citealt{paperI}). This feature  is
a numerical artifact of the finite resolution of the
simulation. Similar errors in the light curves were also present in 
previous simulations for the $k=0$ case
(e.g., our light curve in the case $k=0$ is nearly identical to that 
by \citealt{ZM09} as depicted in  \citealt{paperI}).

A smaller contribution  (although not easily quantifiable)  to the pan-chromatic 
dip in the lightcurve is due to the particular initial conditions chosen 
in this paper. In fact, as the jet initially has  sharp edges (a step function in the $\theta$-direction), once 
the simulation starts there is a relaxation period occurring in the lateral direction 
on a dynamical timescale (as a rarefaction wave propagates from the 
edge of the jet towards its center). This lateral transient phase  triggered by the 
sharp-edged jet is also imprinted  in the lightcurves around the time of the
dip or flattening, and, contrary to the limited resolution artifacts, is not expected to go away
as the resolution is increased. This artifact might be
less pronounced for initial conditions that are smoother in the lateral
direction (e.g., a jet with an initial Gaussian angular profile).

Apart from this early-time, artificial  feature, there is the expected pan-chromatic
jet break that is present at  all frequencies above $\nu_m$ and  is observed  between a day for $k = 2$ to several days for
$k = 0$. These jet break features are  discussed in more detail below.

\subsection{Jet Breaks}

Figure~\ref{fig10} plots the shape of the jet break, i.e. the temporal
decay index of the light curve, $\alpha\equiv- d\log F_\nu/d\log
t_{\rm obs}$, as a function of observer time, $t_{\rm obs}$, for
different observed frequencies and $k$-values. We shall first
discuss  the pan-chromatic jet break features at frequencies that are above the
typical synchrotron frequency at the time of the jet break,
$\nu>\nu_m(t_{\rm obs,j})$.  As shown in Figure~\ref{fig10}, the temporal
decay of the light curve becomes smoother for increasing
$k$, as derived in analytic models ~\citep[][hereafter
KP00]{KP00}. However, the steepening in the lightcurve occurs within a
significantly smaller observed time  period than that
predicted  by analytic models. Most of the increase in $\alpha$ occurs
over a factor of $\approx 3-5$ in time for $k=0$ (compared to a decade in time
predicted  in KP00) and within about one decade in time for $k=2$ (compared to
four decades in time predicted in KP00). The relatively sharper jet break
(compared to analytic expectations) in a stratified medium may permit
the detection of such a jet break. We also note that there is an
``overshoot'' in the value of the 
temporal decay index $\alpha$ just after the jet break, which 
is more prominent for lower $k$-values
~\citep[in agreement with previous results;][]{Granot07}. After this
overshoot $\alpha$ gradually decreases, and there is also a noticeable
curvature in the lightcurve as the flow becomes mildly relativistic
and eventually approaches the Newtonian regime. 
The effects of electron cooling on the shape of the jet break appear 
to be rather modest in most cases.

At low frequencies, $\nu <\nu_m(t_{\rm obs,j})$ (see Figure \ref{fig10}, 
upper panel), there is only a very
modest increase in $\alpha$ near $t_{\rm obs,j}$. On the other hand,
when the break frequency $\nu_m$ sweeps past the observed frequency
$\nu$, a very sharp break is seen (i.e. increase in $\alpha$). Both
features are present for $k=0$, and we find here that they also persist
for higher $k$-values. Moreover, we also find that this break is
sharper for smaller $k$-values. This is because the corresponding
spectral break (at $\nu_m$) is very sharp for our simple broken
power-law spectral emissivity model, and is not degraded  by the
contribution from multiple parts of the jet at smaller $k$-values
(in addition $\nu_m$ decreases somewhat faster in time at $t_{\rm
obs}>t_{\rm obs,j}$ for smaller $k$-values). We expect that a more
realistic synchrotron emissivity function would result in a
significantly smoother spectral break at $\nu_m$,
which would in turn lead to a correspondingly
smoother temporal break.
\\

\begin{figure}
 \includegraphics[width=0.45\textwidth]{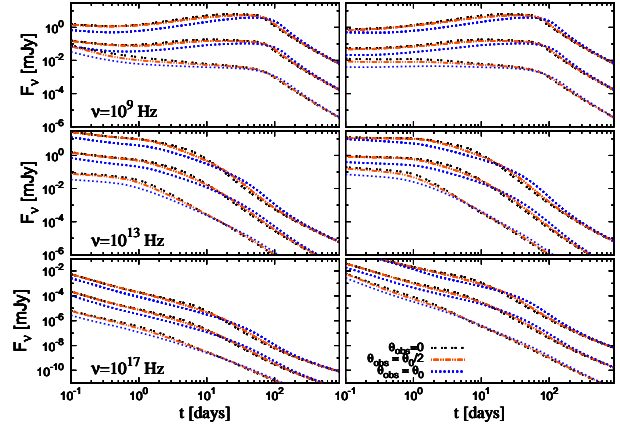}
\caption{
   Light curves corresponding to $\nu = 10^{9}$, $10^{13}$,
   $10^{17}\;$Hz (from top to bottom panels) for different viewing
   angles $\theta_{\rm obs}$ (normalized to the jet initial
   half-opening angle $\theta_0$) and external density profiles ($k =
   0,\,1,\,2$), with (left panels) and without (right panels) electron
   cooling.  The lightcurves corresponding to $k=0$ and $k=1$ are
   multiplied by 1000 and 30, respectively. The lightcurves include
   the contribution from a mapped Blandford-McKee solution (with $20
   \leq \Gamma \leq 200$) and the numerical simulation (with $1 \leq
   \Gamma \leq 20$).}
\label{fig11}
\end{figure}

\begin{figure}
 \includegraphics[width=0.4\textwidth]{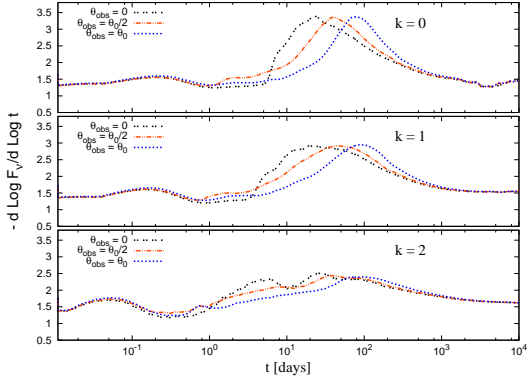}
\caption{
``Shape of the jet break'', i.e. temporal decay of light curve, given
by $\alpha\equiv- d\log F_\nu/d\log t_{\rm obs}$ as a function of
$t_{\rm obs}$, including 
electron cooling, at $\nu = 10^{17}\;$Hz $> \nu_m$.}
\label{fig12}
\end{figure}

Figure~\ref{fig11} shows afterglow lightcurves for three different observed
frequencies ($\nu = 10^{9}$, $10^{13}$, $10^{17}\;$Hz; top to bottom
panels), external density profiles ($k = 0,\,1,\,2$), and viewing
angles ($\theta_{\rm obs}/\theta_0 = 0,\,0.5,\,1$), both with and
without electron cooling (left and right panels, respectively).
Figure~\ref{fig12} shows the corresponding values of the temporal
decay index $\alpha$ for $\nu=10^{17}\;$Hz.  Figures~\ref{fig11} and
\ref{fig12} show that the shape of the jet break is predominantly regulated by the
change in viewing angle (within the initial jet aperture,
$0\leq\theta_{\rm obs}/\theta_0\leq 1$) rather  than by the external
density power-law index $k$ (in the range $0\leq k\leq 2$).  For
$\theta_{\rm obs}=0$ most of the steepening occurs within a factor of
$\sim 2-4$ in time for $k = 0,\,1,\,2$ while for $\theta_{\rm
obs}/\theta_0 \sim 0.5-1$ it takes $\sim 1-2$ decades for $k =
0,\,1,\,2$. This is particularly  interesting because  previous analytical work
have argued that the effect of varying $k$ should be significantly larger.  It can also be seen in Figure~\ref{fig12}
that the jet induced steepening starts earlier and ends later for
larger $k$-values and for larger viewing angles (or $\theta_{\rm
obs}/\theta_0$ values). Also, the overshoot in the value of $\alpha$
is larger for greater $k$-values or $\theta_{\rm obs}/\theta_0$
values. The jet break time is also observed to occurs later for larger viewing angles at
all values of $k$, and varies over a factor of $\sim 3-5$ for $0\leq\theta_{\rm
obs}/\theta_0\leq 1$.

The change in the jet break duration with $k$ is due to the slower
evolution of $\Gamma$ with $t$ or $R\approx ct$ as well as $t_{\rm
obs}$ for larger $k$-values ($\Gamma\propto R^{(k-3)/2}\propto t_{\rm
obs}^{(k-3)/(8-2k)}$ for a spherical flow). For $\theta_{\rm obs} = 0$
and $\nu > \nu_m(t_{\rm obs,j})$, the jet break duration roughly
corresponds to the time it takes the beaming cone to grow past the
limb-brightened outer part of the image. If crudely neglecting lateral spreading 
(since most of the emission near the jet break time is from within the initial jet
aperture, \citealt{2001grba.conf..300P}), so that the dominant
effect is the ``missing emission'' from outside the edges of the jet
\citep{Granot07}), and requiring that the beaming cone (of angle
$\theta\lesssim 1/\Gamma$ around the line of sight) grows by a
factor of $f_k$, then this would correspond to a factor of $\sim
f_k^{(8-2k)/(3-k)}$ in observed time. However, the resulting  image is more
limb-brightened for smaller $k$-values \citep{2001ApJ...551L..63G, Granot08},
and, as a result, one might estimate  $f_{k=0} \sim 1.3$,
$f_{k=1} \sim 1.4$, $f_{k=2}\sim 1.5$, which would result in factors of $\sim
2$, $\sim 3$ and $\sim 5$ in the observed time, in rough agreement
with our numerical results.  

As to the effect of the viewing angle for a fixed value of $k$, the
addition to the duration of the jet break relative to $\theta_{\rm
obs}=0$ corresponds approximately to the time it takes the edge of the beaming
cone ($1/\Gamma$) to grow from $\theta_0$ to $\theta_0+\theta_{\rm
obs}$. Thus, for $\theta_{\rm obs}=\theta_0$ this corresponds to a
factor of $2$ decrease in $\Gamma$, or a factor of $\sim
2^{(8-2k)/(3-k)}$ increase in the observed time (i.e. factors of $\sim
6$, $\sim 8$ and $\sim 16$ for $k = 0$, 1, and 2, respectively). This
is in rough agreement with our numerical results. According to this simple
estimate, the duration of the jet break for $\theta_{\rm obs} =
\theta_0$ and $k = 2$ should be a factor of $\sim
(2f_2)^{(8-2k)/(3-k)} \sim 3^4 \sim 81$ in time, or almost two decades
in observed time, also in agreement with the  results of our calculations.

\subsection{Radio Calorimetry}

\begin{figure}
 \includegraphics[width=0.42\textwidth]{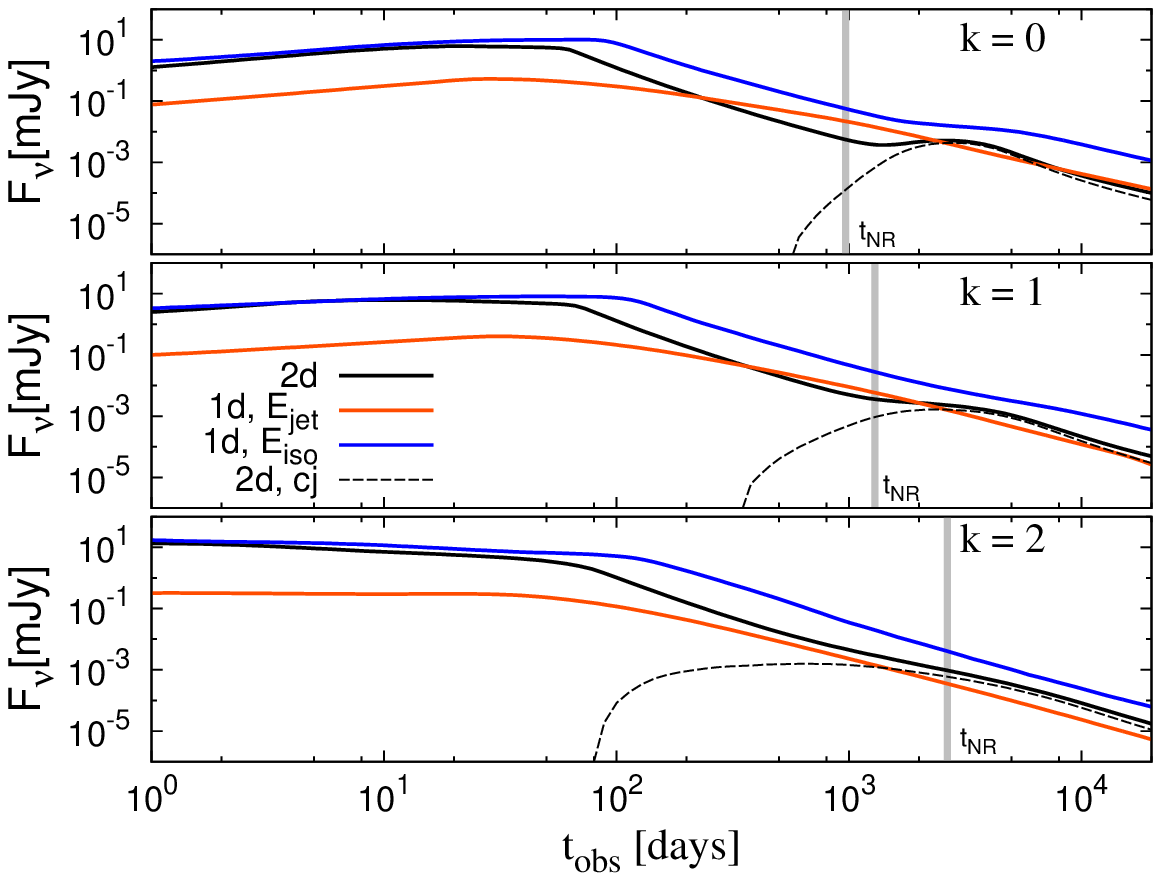}\\
 \includegraphics[width=0.4\textwidth]{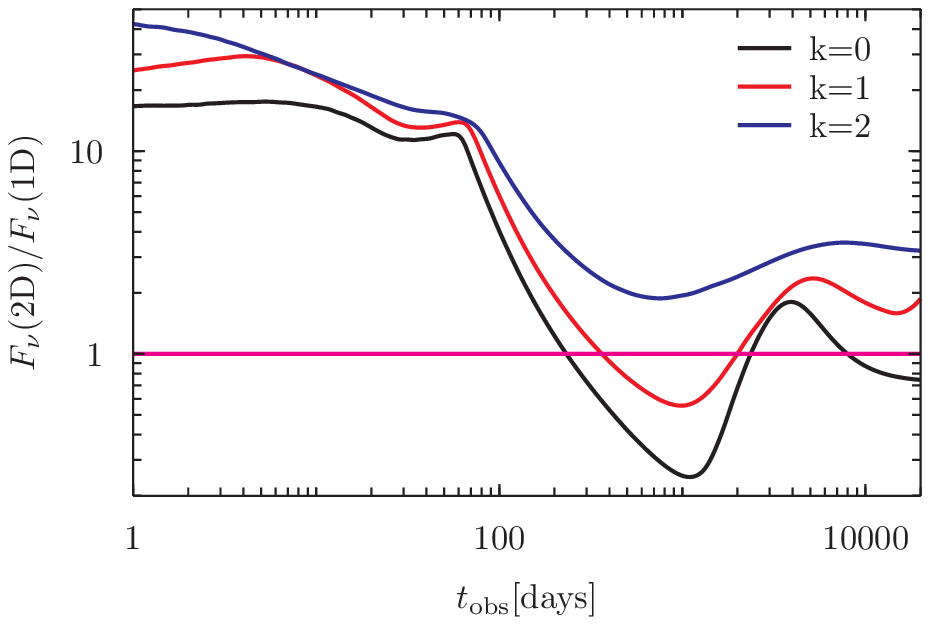}
\caption{Light curve at $\nu = 10^9$~Hz for the 2d runs
($k=0,1,2$), for spherical 1d simulations with $E=E_{\rm jet}$
and for a cone with half-opening angle $\theta_0$
computed from spherical 1d simulations with $E=E_{\rm iso}$.
The contribution due to the counterjet is included 
in the lightcurves, and it is also shown (dotted curves)
for the 2d simulations.
}
\label{fig13}
\end{figure}

Figure~\ref{fig13} shows the radio lightcurves 
(at $\nu=10^9$~Hz) for $k = 0,\;1,\;2$
from our two-dimensional numerical simulations of a double-sided jet, 
as well as for a spherical blast wave with the same true energy and a double-sided
cone of fixed half-opening angle $\theta_0$ calculated  from  a spherical
blast wave with the same isotropic equivalent energy  (where
$\theta_{\rm obs} = 0$ in the two non-spherical cases).
As expected, the lightcurves computed from a spherical blast wave
with the same isotropic energy and from the two dimensional simulation
match reasonably well at early times. 
For the double-sided jet, it can be seen that the bump
in the lightcurve near the non-relativistic transition time, 
caused because the counter-jet (whose contribution is indicated by a
{\it dashed line}) becomes visible,  is much more prominent for low
values of $k$ and becomes significantly  more modest for larger $k$-values. This
effect is caused by the more gradual deceleration of the jet for
larger $k$-values (as the same mass of external medium is swept-up
over a larger range in radii), which causes the counter-jet to become
visible more gradually, resulting in a wider, lower peak flux bump. In particular,
for $k=2$ it amounts to a fairly modest and rather slow flattening of
the lightcurve, which might be hard to discern observationally. This might,
however, not help explain the lack of a clear flattening or
rebrightening in the late radio afterglow of GRB~030329
~\citep[e.g.,][]{2007ApJ...664..411P}, since in that case detailed
afterglow modeling favors a uniform external density
~\citep[$k=0$;][]{vDH08}.

Comparison of the radio flux at late times from a double-sided jet and
from a spherical blast wave with the same true energy near the
non-relativistic transition time shows that they are broadly similar
but may differ by up to a factor $\lesssim 3$. 
For $k = 0$ and $k = 1$ the
spherical analog slightly over-predicts the flux before the
contribution from the counter jet becomes important, and
under-predicts the flux once the emission from the counter-jet becomes
dominant, while for $k = 2$ the spherical analogue consistently 
under-predicts the flux, by up to a factor of $\lesssim 3$.
This may result in an small but not negligible error 
in the estimation of the true energy in the double-side jet assuming
a spherical sub-relativistic flow, as is commonly done in radio 
calorimetry studies
\citep{2007ApJ...654..385K,2003Natur.426..154B,2005ApJ...619..994F,2006ApJ...641L..13G,1998Natur.395..663K},
both over- or under- estimating the real true energy depending on 
the stratification of the ambient medium and the observer time).

\section{Discussion}\label{dis}

We have studied the dynamics of GRB jets during the afterglow stage as  they propagate into different external density profiles,
$\rho_{\rm ext} = Ar^{-k}$ for $k=0,\,1,\,2$, using detailed hydrodynamic simulations. Our main results, which relate both to the dynamics and the resulting afterglow emission,  can be summarized as follows.

For the same initial half-opening angle $\theta_0$ and
external density at the jet break radius (which is defined by
$\Gamma_1(R_j)=\theta_0^{-1}$), the lateral spreading is initially (at
$R<R_j$) larger for higher $k$-values. This arises because  at the same radius (or lab frame time) the typical
Lorentz factor is lower. At late times ($R>R_j$) the situation is reversed, and the effective jet
opening angle at a fixed lab frame time is  similar  for different
$k$-values. Since for higher $k$-values a larger range of radii is
required in order to sweep-up the same amount of mass, the
whole evolution extends over a much wider range of radii and times. As
a result, the jet break in the afterglow lightcurve is smoother and more
gradual, the non-relativistic transition occurs later, and the flow
approaches spherical symmetry more slowly and over longer
timescales. The effective jet opening angle is observed to increase only
logarithmically with lab frame time (or radius) once the jet comes into
lateral causal contact (i.e. when $\Gamma$ drops below
$\theta_0^{-1}$). 

As long as the jet is relativistic, most of the
energy and emission are concentrated near the head of the jet
while the slower material at the
edges carries relatively little energy (even though it carries a substantial
fraction of the swept-up rest-mass).  This holds true for all $k$-values. 
Once the jet becomes sub-relativistic, at $t>t_{\rm NR}(E_{\rm iso})$, it
quickly spreads laterally and swiftly starts to approach spherical symmetry. The energy weighted mean value of $u(t)$ is observed to be of order unity 
at $t/t_j \sim 2$ rather than at $t \sim t_{\rm NR}(E_{\rm iso})$, as one might naively expect.
We find that there is little $k$-dependence on the temporal evolution of $\theta_j$, so that irrespective of the external medium radial profile, all of the expanding jets  approach
spherical symmetry at similar times  ($\sim 1-1.5$ decades after
$t_j$). A similar conclusion can be reached  from the calculated  
evolution of  $R_\parallel/R_\perp$  with  $t/t_j$.

We find that contrary to the expectations of analytic
models, the shape of the jet break is affected more by the viewing
angle (within the initial jet aperture, $0\leq\theta_{\rm
obs}/\theta_0\leq 1$) than by the steepness of the external density profile (for $0\leq
k\leq 2$). Larger viewing angles result in a  later jet break time and
a smoother jet break, extending over a wide range in time, and with
a larger overshoot (initial increase in the temporal decay index
$\alpha$ beyond its asymptotic value), which is observed to be  more prominent for
lower $k$-values. Larger $k$-values result in more gradual jet breaks,
but the sharpness of the jet break is affected even slightly more by
the viewing angle as argued above. The counter-jet becomes visible around $t_{\rm
NR}$, and for $k=0$ this results in a clear bump in the light
curve. However, for larger $k$-values the jet deceleration is more
gradual and as a result  a wider and lower bump is produced, which becomes hard to
detect  for $k=2$, where it reduces to a mild flattening in
the light curve. This may explain the lack of a clear counter-jet
signature in some late time radio afterglow light curves of long
duration GRBs although the dynamical
complexity of their surrounding circumburst medium seriously
limits the validity of a non-evolving  power-law density profile \citep[e.g.][]{2005ApJ...631..435R}. 

Finally, we showed that the use of a spherical blast wave for 
estimating the total energy of the jet, as is commonly done in radio calorimetry 
studies, results in an error in the estimation of the true energy content
of the jet that depends on the stratification of the ambient medium (being 
on average larger for $k=2$). In particular, in the case $k=2$, 
the spherical blast wave analogy consistently overestimates
the true energy, while for the cases $k=0$ and $k=1$ it produces and 
under- or an over-estimate depending on whether the estimation of the jet energy
is done before or after the non-relativistic transition time.

\acknowledgements
We are  grateful to A. MacFadyen, W. Lee and W. Zhang  for discussions.
This research was supported by the David and Lucille Packard Foundation (ERR and FDC),
the NSF (ERR) (AST- 0847563), the ERC advanced research grant ``GRBs'' and a DGAPA
postdoctoral grant from UNAM (DLC).
We aknowledge the support by S. Dong for administrating the Pleiades
supercomputer, maintained and operated by the University of California
at Santa Cruz, where the numerical calculations in this paper were performed.


\end{document}